# Improving Web Search Using Contextual Retrieval

Dilip K. Limbu, Andrew M. Connor, Russel Pears & Stephen G. MacDonell
*Software Engineering Research Lab, Auckland University of Technology*

**Abstract**

*Contextual retrieval is a critical technique for today's search engines in terms of facilitating queries and returning relevant information. This paper reports on the development and evaluation of a system designed to tackle some of the challenges associated with contextual information retrieval from the World Wide Web (WWW). The developed system has been designed with a view to capturing both implicit and explicit user data which is used to develop a personal contextual profile. Such profiles can be shared across multiple users to create a shared contextual knowledge base. These are used to refine search queries and improve both the search results for a user as well as their search experience. An empirical study has been undertaken to evaluate the system against a number of hypotheses. In this paper, results related to one are presented that support the claim that users can find information more readily using the contextual search system.*

*Keywords:* Web searching, information retrieval.

## 1. Introduction

Research in contextual retrieval approaches has become prominent in the interactive Web information retrieval field. The primary goal of contextual retrieval is to acquire a user's information seeking behavior, such as their search activities and responses, and incorporate this information into a search system. The aim is to create a more effective, efficient and personalized interaction employing an appropriate retrieval strategy, by tailoring the system to their preferences. Contextual retrieval has been defined as an information retrieval process that combines search technologies, knowledge about a query, and user context into a single framework in order to provide the most appropriate answer to a user's information need [1].

Despite its growing importance and the development of contextual retrieval approaches, there remains no comprehensive model to fully describe contextual retrieval [2] due to the difficulty of capturing and representing knowledge about users, tasks, and context in a general Web search environment. In effect, contextual retrieval remains as a major long-term challenge [1]. This paper outlines a contextual retrieval system that has been developed to address some of the challenges associated with effective retrieval of information from the WWW.

## 2. Background and Related Work

Previous work in the area of contextual retrieval has focused on three main themes: user profile modeling, query expansion, and relevance feedback. These themes have some characteristics in common, but also many differences. A brief summary of these themes, presented fully elsewhere [3, 4, 5] is useful in order to place the current research in the context of other studies.

### 2.1. Information Retrieval Techniques

**2.1.1. User profile modeling.** Several Web IR systems have explored various user modeling approaches to improve the personalization of a users' Web search experience. A review of these user modeling approaches reveals that they all utilize either user behavior or user preferences to construct a contextual profile. However, none of the approaches considered use a combination of user behavior and preferences. Apart from InfoFinder [6], none of approaches reviewed discusses Boolean query expansions using any form of user contextual profile. Similarly, apart from WebMate [7], these approaches do not have the capability to share a user's contextual profile information with other users, thereby potentially leading to suboptimal performance when the user needs access to information outside their original context. The use of shared contextual profiles or collaborative filtering leverages the collective profiles of a number of users and can assist users with similar interests [8].

While showing promise, prior IR approaches employing user profile modeling have had limited success. Fundamental challenges remain, specifically: i) how to acquire, maintain and represent accurate information about a user's multiple interests with minimal intervention; ii) how to use this acquired information about the user to deliver personalized search results, and iii) how to use information acquired from various users as a knowledge base in large communities or groups?

**2.1.2. Query expansion.** In general, a query expansion approach attempts to expand the original search query by adding further, new or related terms. These terms are added to an existing query either by the user, known as interactive query expansion (IQE), or by the retrieval system, known as automatic query expansion (AQE). These additional terms are intended to increase the accuracy of the search, though comparative studies of the two approaches have produced inconclusive findings regarding the relative merits of each [9, 10]. Our review of query expansion approaches has highlighted a number of alternatives for identifying terms to add to the original query. Of these, the thesauri and concept based approaches are attractive and promising. Approaches utilizing search histories and query logs are also useful, as log data can be used to train parameters to obtain the query expansion terms, based on selected terms from past user searches or queries that are associated with documents in the collection.

The main challenges for current query expansion techniques are: i) which terms should be included in the query expansion; ii) how these terms are ranked or selected; and iii) which levels of query reformulation should be automatic, interactive or manual [11].

**2.1.3. Relevance feedback.** The idea behind relevance feedback (RF) is to take the results that are initially returned from a query and to use information about whether or not those results are relevant to perform a new query. Relevance feedback provides a means for automatically reformulating a query to more accurately reflect a user's interests [12]. RF has been researched extensively in interactive settings and can be exploited using either explicit or implicit feedback. The feedback information can be used to construct a user's contextual profile which is used to query, filter and return relevant information. Despite considerable research, these approaches have not been successfully implemented in Web-based information retrieval [13].

The main challenges faced by current RF mechanisms are: i) how to capture a user's information-seeking behavior and their preferences and to structure this information in such a way as to be able to define a search context that can be refined over time; ii) how to help the user form or join communities of interest while respecting their personal privacy; and iii) how to develop algorithms that combine multiple types of information to compute recommendations.

## 3. Contextual search system

This section provides a brief overview of contextual search system [3, 4, 5]. The contextual search system includes a range of features, such as adaptation of a user's information seeking behavior, recognition of a user's preferences and interests, recommendation of terms, generation of Boolean query and presentation of ranked contextual search results to attempt to improve the user experience. The system uses a three tier architecture, with the core functions being in the contextual search layer.

The contextual search layer links two other layers: presentation layer and database layer. For example, the layer processes requests from the presentation layer (e.g., a user registration) and sends instructions to the database layer to store or retrieve a piece of data (e.g., the registration data). It is the performance of the components in this layer that will affect the ability of the system to enhance the user's web search experience, and it is the components in this layer that are evaluated in this paper. Figure 1 shows the contextual search system architecture.

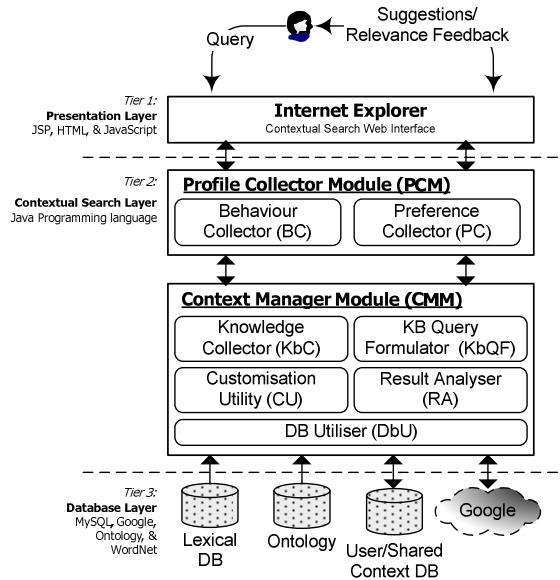

**Figure 1. Contextual search system architecture**

The layer comprises two main modules: the Profile Collector Module (PCM) and Context Manager Module (CMM) to perform the following functions;
1. Gather the user's implicit data, such as entered search queries, visited URLs and meta-keywords.
2. Capture the user's explicit data, such as alternative terms, meta keywords or similar phrases and concepts. This data is sourced from a lexical database, a shared contextual knowledge base (SCKB) and domain-specific ontologies.
3. Construct the user's personal contextual profile and a shared contextual knowledge base using data from step 1 and step 2.
4. Modify the user's initial query to more accurately reflect the user's interests.

Each module consists of several components that perform these various functions, with the PCM components forming the core of the system, as a result this paper focuses on the PCM and a full discussion of the CMM is outside the scope of this paper.

### 3.1. Profile collector module

The PCM is implemented to capture both a user's behavior and preferences as a user's personal contextual profile and structure this information in such a way as to be able to define a search context that can be refined over time. Figure 2 illustrates the functionality of the PCM, a hybrid contextual user profiling approach that captures a user's adaptive search behavior by monitoring and capturing their explicit (i.e., explicit rankings, inputs, and instructions) and implicit (i.e., browsing and typing) data. The PCM constantly acquires and maintains this data with minimal intervention.

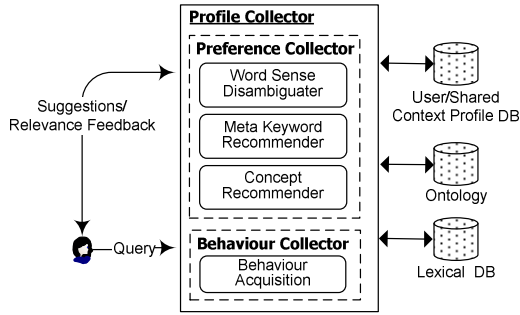

**Figure 2. The profile collector module functionality**

The PCM consists of two specialized components: Preference Collector (PC) and Behavior Collector (BC). Both components gather information seeking behavior from users. A few assumptions were made during the development of the PCM related to privacy concerns and barriers to adoption. It suffices to summarize them into the simple statement that it is assumed that user will wish to share their data if there is perceived benefit in doing so.

Figure 3 provides an overview of the functionality of the PC component, consisting of the Word Sense Disambiguater (WSD), Meta Keyword Recommender (MKR) and Concept Recommender (CR) processes. The PC component learns a user's specific information needs by capturing their explicit preferences and at the same time recommends terms, phrases and concepts that will be of potential interest to the user.

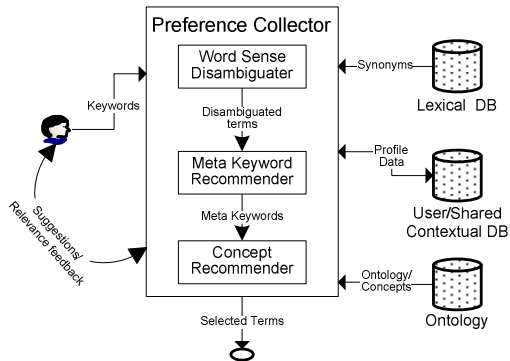

**Figure 3. The PC component functionality**

The PC component employs the nearest-neighbor algorithm to learn a user's specific information needs and to recommend alternative terms. The component uses a two step 'divide and conquer' algorithm to address the scalability of the nearest neighbor. The PC component also employs the relevance feedback approach to support the iterative development of a query by recommending alternative terms for query formulation [4]. However, the effectiveness of learning each user's specific information needs and alternative terms recommendation is directly proportionate to the availability and size of the user's personal contextual profile and shared contextual knowledge base and so the issue of cold start remains.

The PC component starts with the WSD process by accepting a user's search query which may contain one or more keywords. WordNet is used to disambiguate each search keyword. As a result, each query may contain one or more disambiguated terms. WordNet has been used as a word sense disambiguation tool in queries [14] as well as many other applications. Each disambiguated term may contain one or more words. The WSD process removes stop words, repeated/similar words, and similar-to-search-query keywords from each disambiguated term. The process constructs an optimized disambiguated terms vector and then uses a recommendation process to compute the nearest-neighbor in the user's personal contextual profile and the shared contextual knowledge base. This is used to recommend disambiguated terms that are relevant to the user's search query. The user has the option to alternatively select more relevant disambiguated terms that best describe the subject of their query. The MKR process then utilizes the search query, the user's context and the user-selected disambiguated terms to compute and recommend meta keywords. The CR process subsequently takes the search query, the user's context, the selected disambiguated terms and the user-selected meta keywords to compute and recommend concepts based on domain-specific ontologies. Figure 4 provides a summarized depiction of the functionality of the BC component, centered on a Behavior Acquisition (BA) process. The BA process monitors and captures a user's daily Internet search activities to represent a user's behavior.

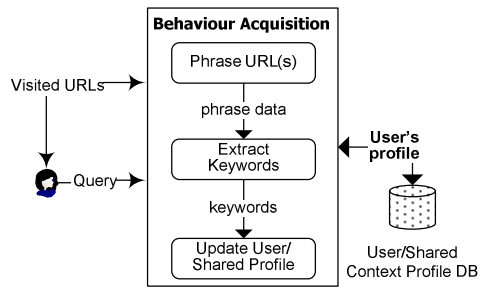

**Figure 4. The BA component functionality**

For example, a user submits a search query and clicks on the URL(s) that are relevant to the query. The BA process computes and extracts a set of meta keywords from each URL. The process also removes stop words, repeated/similar words, and similar-to–search-query keywords from the extracted meta keywords. Each meta keyword may contain one or more words. The process extracts only the first five words from each meta keyword to construct a meta keyword vector. The main reason for removing all further information, limiting the extraction to five words per meta keyword and the presentation of the five meta keywords is to limit the number of keywords/terms in Boolean query (for performance reasons) and to reduce clutter in the user interface. Finally, the BC component stores all information as a user's behavioral data for future use.

## 4. Hypothesis

The study was designed to investigate the performance of the contextual retrieval system along the usability dimensions of effectiveness, efficiency and subjective satisfaction. As described, three aspects to the user study were carried out intended to address five hypotheses. Under the mixed-method approach adopted here, the first hypothesis is essentially addressed using quantitative methods, whilst the remaining hypotheses are essentially qualitative. This paper limits itself to discussing the quantitative results related to the primary hypothesis.

The primary hypothesis examines the subjects' overall information seeking behavior and differences in performance between the contextual search system and the contemporary search engine. This hypothesis is given as;

- **Hypothesis 1 - Find Information Readily:** The contextual retrieval system enables subjects to find relevant information more readily than a standard search engine using their personal profile and shared contextual knowledge base.

The primary hypothesis is further divided into a number of sub-hypotheses to facilitate the capture and analysis of data. Each sub-hypothesis is worded as a null-hypothesis. To test hypothesis 1, five performance aspects (representing effectiveness and efficiency) of the contextual search system and the contemporary search engine were compared;

- **Hypothesis 1.1 - Number of queries:** There is no difference in number of queries entered to reach the target information in the three phases of the study.
- **Hypothesis 1.2 - Number of clicks:** There is no difference in number of clicks clicked to reach the target information in the three phases of the study.
- **Hypothesis 1.3 - Number of hits:** There is no difference in number of hits browsed to reach the target information in the three phases of the study.
- **Hypothesis 1.4 - Number of URLs:** There is no difference in number of URLs visited to reach the target information in the three phases of the study.
- **Hypothesis 1.5 - Length of time:** There is no difference in length of time taken to reach the target information in the three phases of the study.

## 5. Evaluation methodology

To assess the effectiveness of our system we have employed a mixed methods research approach, based on simulated work task situations, questionnaires, and observations that has been informed by earlier theories of cognitive and information-seeking behavior. A qualitative study was conducted to investigate in depth subjects' information seeking behavior for a specific situation. A quantitative study was conducted in parallel to determine the performance of the system along the usability dimensions of effectiveness and efficiency.

The main objective of the user study was to measure the performance of the system along the usability dimensions of effectiveness, efficiency and subjective satisfaction when compared to a contemporary search engine. The empirical study comprised three phases OS-I, OS-II, and OS-III.

The OS-I and OS-II phases were carried out with differing objectives. The aim of the OS-I phase was to determine whether the contextual retrieval system enabled subjects to find relevant information more readily than a standard search engine using their personal contextual profiles. During the OS-I phase, subjects performed six search tasks using the system, a technique similar to that used in other studies in this area [15]. Users had their search behaviors and preferences captured in order to create their personal contextual profiles and to populate a Shared Contextual Knowledge Base (SCKB). However, the SCKB was not accessible to them during their search.

Once the OS-I phase was complete, a second group of subjects repeated the same six allotted search tasks; however these subjects had the SCKB enabled. As such, the aim of the OS-II phase was to determine whether the system enabled subjects to find relevant information more readily than a standard search engine using their personal contextual profiles and the SCKB. This allowed us to measure the contribution of the shared profile to search quality, by comparing the "speed" with which subjects could find data with the first group who did not have access to the shared profile. The OS-III phase was carried out on a contemporary search engine, i.e., Google, to tackle the same six search tasks and to provide a benchmark set of results.

A total of 30 subjects, with different levels of search experience, participated in the three phases. Subjects were randomly assigned to one of these phases, so that there were ten subjects in each group. Before the actual user study, subjects were given the same general instructions, video demonstration, and filled in an entry questionnaire requesting information about their characteristics and search experience. Subjects in the OS-I and OS-II groups filled in a post observation questionnaire so that we could capture their overall reactions to the contextual search system.

## 6. Results

The results presented in this section are from the observation data. The observation data, such as number of queries, number of clicks, number of hits, number of URLs and length of time taken to reach target information, were extracted from observation video recordings, observation notes and the system logs.

Both parametric and nonparametric statistical methods were used to test statistical significance between empirical data. The observational data were interval in nature and parametric methods are thus more appropriate. Both graphical (i.e., Q-Q and Detrended Q-Q plots) and numerical normality (i.e., Shapiro-Wilk) tests were carried out on these interval scale data, to determine whether or not the data were normally distributed. The parametric one-way analysis of variance (ANOVA) and Dunnett's one-tailed tests were performed on those data normally distributed to test statistical significance. In addition, to guard against the possibility that the assumption of normal distribution did not hold, nonparametric Kruskal-Wallis tests were performed. In addition, where appropriate, Tukey's honestly significant difference post hoc tests were used to reduce the likelihood of Type I errors (i.e., rejecting null hypotheses that are true). The significance threshold for all tests was set at $p = 0.05$.

Table 1 shows the summary of the statistical analysis of hypothesis 1 (i.e., find information readily) for all six search tasks, which measured the effectiveness and efficiency of the contextual retrieval system (i.e., OS-I and OS-II) when compared to a contemporary search engine (i.e., OS-III). The results demonstrate that the contextual search impact is significant (actual p-values bolded) in terms of the number of hits browsed (H 1.3) and number of URLs visited (H 1.4) for overall search task completion.

**Table 1. Statistical analysis of H1.1 to H1.5**

| Hypothesis | Parametric | Non-Parametric | Post Hoc Tests |
|---|---|---|---|
| H 1.1 | - | $X^2 = 3.056$, KW $= 0.217$ | - |
| H 1.2 | A = 0.413<br>Dt-ac = 0.474<br>Dt-bc = 0.162 | $X^2 = 0.954$, KW $= 0.621$ | - |
| **H 1.3** | A = **0.005**<br>Dt-ac = 0.163<br>Dt-bc = **0.001** | $X^2 = 10.448$, KW $= 0.005$ | TH–bc = **0.004** |
| **H 1.4** | A = **0.004**<br>Dt-ac = 0.118<br>Dt-bc = **0.001** | $X^2 = 10.648$, KW $= 0.005$ | TH–bc = **0.001** |
| H 1.5 | A = 0.242<br>Dt-ac = 0.559<br>Dt-bc = 0.102 | $X^2 = 3.510$, KW $= 0.173$ | - |
| **Legend** | | | |
| One-way ANOVA P-value | = | A | |
| Dunnett's one-tailed P-value | = | Dt | |
| Kruskal-Wallis P-value | = | KW | |
| Tukey's honestly P-value | = | TH | |
| -ac | = | between OS-I & OS-II | |
| -bc | = | between OS-II & OS-III | |

During the observation study, it was discovered that one of the six assigned tasks was somewhat tricky. Subjects' answers to this task were often imprecise, and subjects commonly took long periods of time, entered more queries and browsed more hits to discover an answer.

For this reason, further detailed one-way ANOVA and Dunnett's one-tailed tests of the non-significant hypotheses were performed for individual search tasks. The results, shown in Table 4, indicate that the contextual search impact is significant in terms of the number of queries (H 1.1), number of clicks (H 1.2), and length of time (H 1.5) for search task six. No significant differences were found for the remaining search tasks.

**Table 2. Analysis of insignificant hypotheses**

| Hypothesis | Parametric | Non-Parametric | Post Hoc Tests |
|---|---|---|---|
| H 1.1 | - | $X^2 = 7.009$, KW $= 0.030$<br>U=17.00, MU-bc $= 0.011$ | - |
| H 1.2 | A = 0.041<br>Dt-ac = 0.145<br>Dt-bc = **0.011** | $X^2 = 8.987$, KW $= 0.011$ | TH–**bc** = **0.032** |
| H 1.5 | A = 0.011<br>Dt-ac = 0.056<br>Dt-bc = **0.003** | $X^2 = 9.177$, KW $= 0.010$ | TH–**bc** = **0.009** |
| **Legend** | | | |
| Mann-Whitney U | = | MU | |

The overall finding of our empirical study (for hypothesis H1) is that the contextual retrieval system delivered either equivalent or improved the Web search effectiveness of Web searches, as subjects actually entered fewer queries to reach the target information in comparison to the contemporary search engine. Similarly, efficiency was improved as subjects browsed fewer hits, visited fewer URLs, clicked fewer clicks and took less time to reach the target information when compared to the contemporary search engine. Given that the contextual retrieval system incorporates additional capabilities (in terms of word sense disambiguation, term recommendation and so on), the fact that there is no additional performance overhead is a promising result.

A promising characteristic of the contextual retrieval system is that it does not force subjects to use its contextual features, nor does it interfere radically beyond the scope of their normal search activities.

In summary, these results provide some evidence to suggest that when the contextual profile and the shared contextual knowledge base are used, the contextual retrieval system improves subjects' overall ability to find information readily.

## 7. Conclusions

This paper has presented research regarding the implementation and evaluation of a contextual retrieval system. The system utilizes a contextual user profile employing both implicit and explicit data to provide relevant information to users that potentially satisfies their information needs.

An observational study has been carried out and initial analysis of the data has shown that the system improves both search effectiveness and efficiency. This study is just one step in this direction. The results of this study serve as a partial view of the phenomenon, and the results could also be interpreted in some other ways. More research needs to be done in order to validate or invalidate these findings, using larger samples, and if possible in a real-life scenario.

## 8. References

[1] Allan, J. et al. (2003) Challenges in information retrieval and language modeling: report of a workshop held at the center for intelligent information retrieval. ACM SIGIR Forum, 37, 31-47.

[2] Wen, J.R., N. Lao, and W.-Y. Ma. Probabilistic model for contextual retrieval. Paper presented at the Annual ACM Conference on Research and Development in Information Retrieval. 2004. Sheffield, United Kingdom.

[3] Limbu, D.K., R. Pears, A.M. Connor and S.G. MacDonell. Contextual relevance feedback in web information retrieval, Paper presented at the 1st International Symposium on Information Interaction in Context, 2006. Copenhagen, Denmark.

[4] Limbu, D.K., R. Pears, A.M. Connor and S.G. MacDonell. Contextual and Concept-Based Interactive Query Expansion. Paper presented at the 19th Annual Conference of the National Advisory Committee on Computing Qualifications, 2006. Wellington, New Zealand.

[5] Limbu, D.K., R. Pears, A.M. Connor and S.G. MacDonell. A Framework for Contextual Information Retrieval from the WWW. Paper presented at the 14th International Conference on Intelligent and Adaptive Systems and Software Engineering, 2005, 185-189.

[6] Krulwich, B. and C. Burkey, The InfoFinder agent: learning user interests through heuristic phrase extraction. IEEE Expert, 1997. 12(5): p. 22 - 27.

[7] Chen, L. and K. Sycara. WebMate: A Personal Agent for Browsing and Searching. Paper presented at the International Conference on Autonomous Agents. 1998. Minneapolis, Minnesota, United States.

[8] Pennock, D.M. and E. Horvitz. Collaborative filtering by personality diagnosis: A hybrid memory- and model-based approach. Paper presented at the IJCAI Workshop on Machine Learning for Information Filtering. 1999. Stockholm, Sweden.

[9] Koenemann, J. and N.J. Belkin. A case for interaction: a study of interactive information retrieval behavior and effectiveness. Paper presented at the SIGCHI conference on Human factors in computing systems: common ground 1996. Vancouver, British Columbia, Canada.

[10] Beaulieu, M., Experiments on interfaces to support query expansion. Journal of Documentation, 1997. 53(1): p. 8-19.

[11] Bates, M.J., Where should the person stop and the information search interface start? Information Processing & Management, 1990. 26(5): p. 575-591.

[12] Allan, J. Incremental relevance feedback for information filtering. Paper presented at the 19th annual international ACM SIGIR conference on Research and development in information retrieval. 1996. Zurich, Switzerland.

[13] Croft, W.B., S. Cronen-Townsend, and V. Lavrenko. Relevance Feedback and Personalization: A Language Modeling Perspective. Paper presented at the DELOS Workshop: Personalisation and Recommender Systems in Digital Libraries. 2001. Dublin City University, Ireland.

[14] Liu, S., Yu, C. & Meng, W. Word sense disambiguation in queries. Paper presented at the 14th ACM international conference on Information and knowledge management, 2005. Bremen, Germany.

[15] Borlund, P., The IIR evaluation model: a framework for evaluation of interactive information retrieval systems. Information Research, 2003, 8(3): p. 1-38.